\documentclass[prb,superscriptaddress,twocolumn,showpacs]{revtex4-1}

\usepackage{graphicx}
\usepackage{amssymb,amsmath}
\usepackage{dcolumn}
\usepackage{bm}
\usepackage{color,soul}
\usepackage{enumerate}
\usepackage{float}

\begin{document}

\title{Information processing with topologically protected vortex memories in exciton-polariton condensates}

\author{H. Sigurdsson}
\affiliation{Division of Physics and Applied Physics, Nanyang
Technological University 637371, Singapore} \affiliation{Science
Institute, University of Iceland, Dunhagi-3, IS-107, Reykjavik,
Iceland}

\author{O. A. Egorov}
\affiliation{Institute of Condensed Matter Theory and Solid State Optics, Abbe
Center of Photonics, Friedrich-Schiller-Universit\"{a}t Jena, Max-Wien-Platz
1, 07743 Jena, Germany}

\author{X. Ma}
\affiliation{Institute of Condensed Matter Theory and Solid State Optics, Abbe
Center of Photonics, Friedrich-Schiller-Universit\"{a}t Jena, Max-Wien-Platz
1, 07743 Jena, Germany}

\author{I. A. Shelykh}
\affiliation{Division of Physics and Applied Physics, Nanyang Technological University 637371, Singapore}
\affiliation{Science Institute, University of Iceland, Dunhagi-3, IS-107, Reykjavik, Iceland}

\author{T. C. H. Liew}
\affiliation{Division of Physics and Applied Physics, Nanyang Technological University 637371, Singapore}

\date{\today}

\begin{abstract}
We show that in a non-equilibrium system of an exciton-polariton condensate, where polaritons are generated from incoherent pumping, a ring-shaped pump allows for stationary vortex memory elements of topological charge $m = 1$ or $m = -1$. Using simple potential guides we can choose whether to copy the same charge or invert it onto another spatially separate ring pump. Such manipulation of binary information opens the possibility of a new type of processing using vortices as topologically protected memory components.
\end{abstract}

\maketitle

\section{Introduction}
The exciton-polariton Bose-Einstein condensate~\cite{Kasprzak2006} (BEC) in planar microcavities has become subject to much research and study~\cite{Microcavities2007} in the past decade with the promise of polariton condensation happening at much higher temperature than atomic condensates.~\cite{Christopoulos2007, Sun2010} BECs are low temperature systems of integer spin bosons in which, below a certain critical temperature, a large fraction condense to the ground state and become macroscopically coherent.~\cite{ReviewBEC} Being supercooled quantum systems, they offer a wide variety of topological phases and excitations such as the quantum vortex which will be the main focus in this paper.

The exciton-polariton is a quasiparticle formed by the strong coupling of a microcavity photon mode to electronic excitations in quantum wells embedded in a microcavity. The most notable features of this composite bosonic particle are its light effective mass (about 4 to 5 orders smaller than the electron mass) which arises from the photonic component, and strong binary interactions from the excitonic one. Polaritons also have a short lifetime (few orders of picoseconds), which inhibits thermalization to the lattice temperature. However, polariton-polariton scattering processes allow polaritons to relax fast enough into a macroscopically occupied ground state, effectively creating a polariton condensate.~\cite{Imamoglu1995}

The quantum vortex is one of the most well studied topological defects in atomic BECs and has been observed in both polariton parametric oscillators~\cite{Krizhanovskii2010, Tosi2011, Guda2013} and non-resonantly excited polariton BECs.~\cite{Lagoudakis2008, Lagoudakis2009} These defects consist of a vortex core, where the condensate density reaches its minimum and phase becomes singular, and a circulating superfluid flow around, with phase winding being an integer number of $2\pi$.~\cite{ReviewBEC} Its topological stability makes the vortex a prime candidate for a robust binary memory component, and recent works have considered control of the path of moving vortices.~\cite{Pigeon2011,Sanvitto2011,Flayac2012,Solnyshkov2012,Pavlovic2012}

In this paper we show that stable vortex solutions of charge $m = \pm1$ are supported by an incoherent ring-shaped pump. Such pump shapes were considered experimentally, demonstrating pattern formation,~\cite{Manni2011, Cristofolini2013} evaporative cooling~\cite{Askitopoulos2013} and vortex-antivortex arrays.~\cite{Manni2013} Unlike in cases of deterministic vortex generation,~\cite{Liew2007,Liew2008,Boulier2014} the vortices that we predict represent multistability in the system and are challenging to observe experimentally due to the random selection of the vortex sign during condensation, which is averaged to zero in multishot experiments. Nevertheless, one can expect that coherent pulses can deterministically select the vortex sign allowing their detection.~\cite{Marchetti2010} Alternatively, it was recently shown that very high quality microcavities can be used to make single shot measurements, allowing vortex detection in ring-shaped traps.~\cite{Liu2014} Vortex states can be read by using interferometry,~\cite{Lagoudakis2008} while other methods have been discussed in Ref.~[\onlinecite{Wouters2009}], such as detection of the wavevector of polaritons around the vortex core.

\begin{figure}
\centering
\centering
   \includegraphics[width=\linewidth]{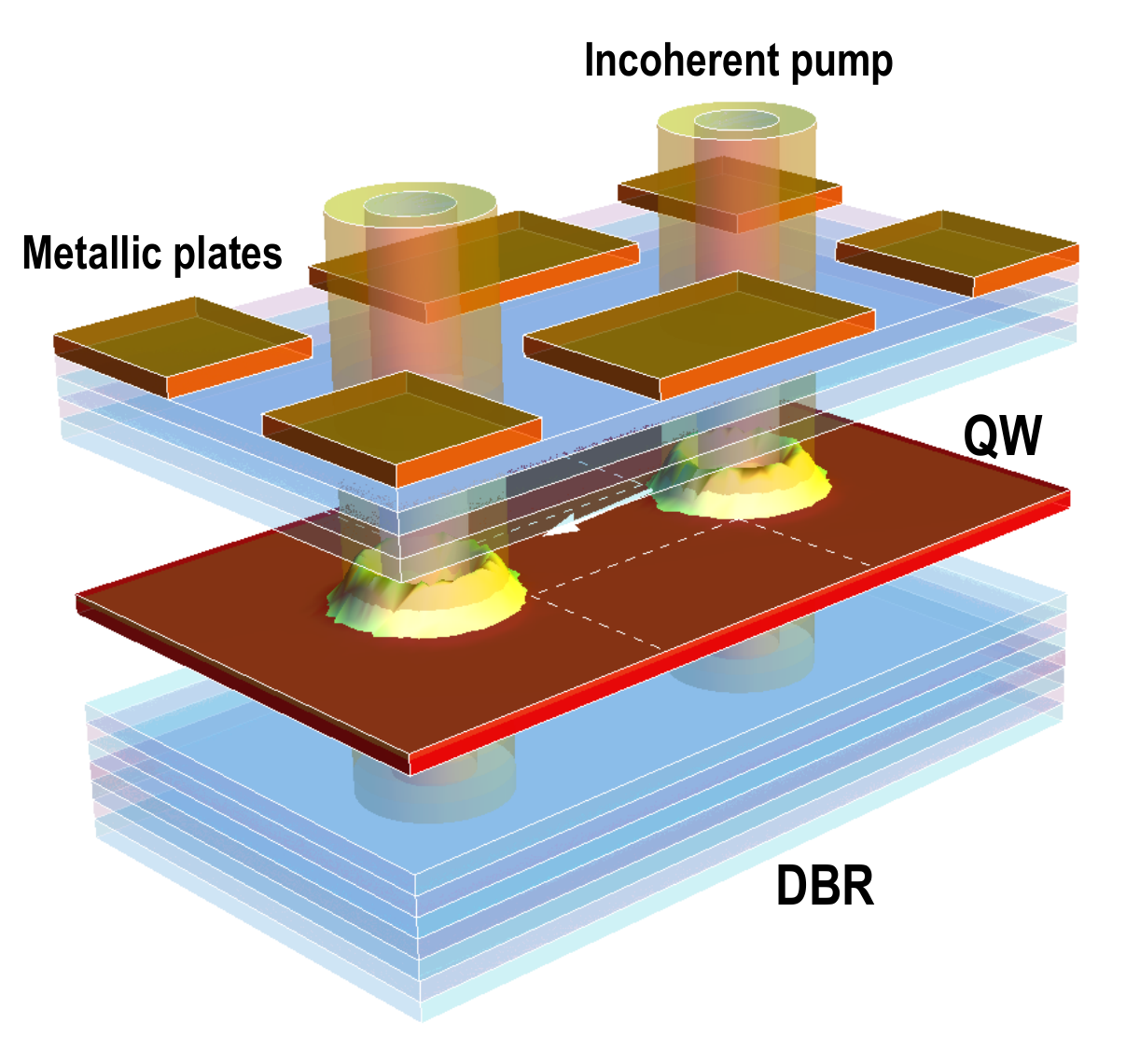}

\caption{(Color online) A schematic showing the exciton-polariton microcavity composed of $\lambda/2$ AlAs cavity with a GaAs quantum well (QW) inside. Distributed Bragg reflectors (DBR) localize the photonic field within the cavity. The ring-shaped incoherent optical pumps create an exciton reservoir in the QW which in turn generates polaritons which form a vortex state. A grid like pattern of metallic plates helps to guide the polaritons from one pump to another.}
\label{fig1}
\end{figure}

Using simple potential guides shown in Fig.~\ref{fig1}, we find that a vortex charge can be both copied or inverted to a second spatially separate ring pump even under a large amount of stochastic noise. This satisfies the standards of copying binary memory, with high fidelity, and of a NOT gate. The choice of whether transferring the same charge $(\pm1 \to \pm1)$ or the inverse charge $(\pm1 \to \mp1)$ can be controlled by either changing the length scales of the potential guide or the distance between the ring pumps.

\section{Theoretical Model}
Using mean field theory and assuming the spontaneous formation of the exciton-polariton condensate, an open-dissipative Gross-Pitaevskii (GP) model describes our incoherently pumped condensate coupled with an exciton reservoir. The polariton order parameter $\Psi$ is described by a GP-type equation and the exciton reservoir density $n_R$ by a rate equation.~\cite{Wouters2007}
\begin{align} \label{eq.polar} \notag
i\hbar \frac{\partial \Psi}{\partial t} &= \Big[ - \frac{\hbar^2}{2m} \nabla_{\perp}^2 + V(\mathbf{r}) + g_c|\Psi|^2  + g_R n_R(\mathbf{r},t) \\
& + i \frac{\hbar}{2}(R n_R(\mathbf{r},t) - \gamma_c)  \Big] \Psi + P_c(\mathbf{r},t)
\end{align}
\begin{equation} \label{eq.reserv}
\frac{\partial n_R}{\partial t} = -(\gamma_R + R |\Psi|^2 ) n_R(\mathbf{r},t) + P_R(\mathbf{r}).
\end{equation}
Here the kinetic energy of polaritons is characterized by effective mass, $m$. $V(\mathbf{r})$ represents any potential patterning of the microcavity, which can be achieved by a variety of techniques, such as: reactive ion etching,~\cite{Bloch1997,Wertz2010,Gao2012} mirror thickness variation,~\cite{Kaitouni2006} stress application,~\cite{Balili2007} metal surface deposition~\cite{Lai2007,Kim2013} or optical means.~\cite{Amo2010} The constants $g_c$ and $g_R$ characterize the strengths of polariton-polariton and polariton-reservoir interactions, respectively. $R$ defines the condensation rate, while $\gamma_c$ and $\gamma_R$ represent the decay rates of polaritons and reservoir excitons, respectively. We allow for both incoherent (non-resonant) pumping, $P_R(\mathbf{r})$, as well as coherent (resonant) pumping $P_c(\mathbf{r},t)$.

\begin{figure}
    \centering
    \includegraphics[width=\linewidth]{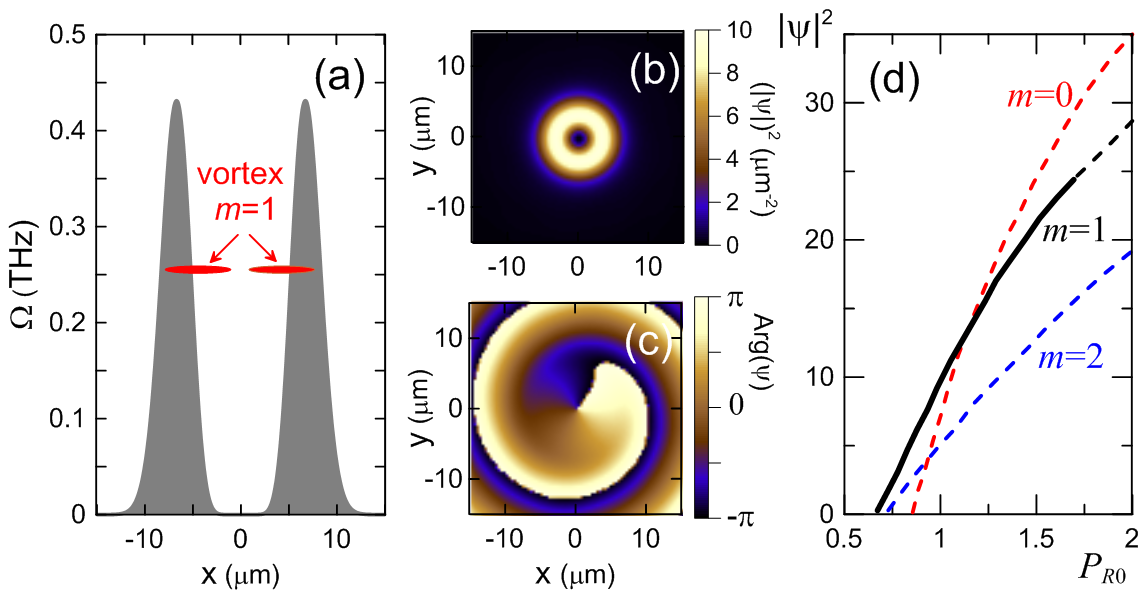}
\caption{(Color online) (a) The gray curves show the frequency shift (blue shift), $\Omega(r)$, induced by the exciton reservoir created by the ring shaped incoherent pump. Polaritons typically condense inside the effective trap potential represented by $\Omega(r)$, with frequency marked in red. (b) Density profile $|\Psi|^2$ of the vortex with charge $m=1$ and its phase profile (c) for $P_{R0}=1$ ps$^{-1}$ $\mu$m$^{-2}$. (d) Bifurcation diagrams of different states supported by the ring pump. Maxima of the polariton density $|\Psi|^2$ vs pump amplitude $P_{R0}$. Dashed lines represent unstable solutions.}
\label{fig2}
\end{figure}

\section{Bistability of vortices under ring shaped pumping}
Let us first consider the case of a ring-shaped incoherent pump, in the absence of any potential ($V(\mathbf{r})=0$). The ring pump profile can be described by the function type:
\begin{equation} \label{eq.pump}
P_R(r) = P_{R0} \left( \frac{r}{w_{1}} \right)^{10} e^{-(r/w_{2})^2},
\end{equation}
where $r = \sqrt{x^2 + y^2}$. Owing to the many particle interaction effects the incoherent pump induces a blue shift in the frequencies of the polaritons $\Omega(r)=g_R n_R/\hbar=g_R P_R(r)/(\hbar \gamma_R)$, shown in Fig.~\ref{fig2}a. It is found that the ring-shaped pump supports a stable vortex solution of charge $m = \pm 1$, provided that the pump intensity overcomes some threshold value (Figs.~\ref{fig2}b and 2c). In general, among these fundamental vortices (with charges $m = \pm 1$) the ring-shaped pump also supports the solutions with other charges including a non-vortex state with $m = 0$ (Fig.~\ref{fig2}d). The latter one is characterized by the formation of the polariton condensate in the center of the ring pump. Similar solutions have been observed recently in Refs.~[\onlinecite{Cristofolini2013, Askitopoulos2013}]. A standard linear stability analysis has been performed to prove the stability of the solutions. It turned out that non-vortex solutions as well as vortices with the charge $m= \pm 2$ are unstable for our particular parameters of the ring pump (this is in contrast to the stability of $m=\pm2$ vortices injected in polariton parametric oscillators with small momentum~\cite{Sanvitto2010, Szymanska2010}). The fundamental vortices with charges $m = \pm 1$ are the only stable solutions and experience destabilization for very high pumping rate (for $P_{R0}\gtrsim1.7$ in Fig.~\ref{fig2}d). The vortex solutions with $m=+1$ and $m=-1$ are equivalent and can be used for bistability schemes. Note that the ring shaped pumping is essential; Gaussian shaped pumps are known not to support stable vortices in the steady state.~\cite{Ostrovskaya2012} The ring shaped pumping also negates the need for an additional hard parabolic trapping, where vortex solutions are also known to exist.~\cite{Keeling2008, Rodrigues2014}

\section{Generating single vortex states}
We now introduce $V(\mathbf{r})$, describing the potential guides which can be achieved using metallic layers, as shown in Fig. \ref{fig1}, which blueshift the polaritons. The parameters chosen for our system correspond to the experimental results of [\onlinecite{Roumpos2011}]. The polariton mass is set to $m = 10^{-4} m_e$ where $m_e$ is the free electron mass. The decay rates are chosen as $\gamma_c = 0.033$ ps$^{-1}$ and $\gamma_R = 1.5 \gamma_c$. The interaction strengths are set to $g_c = 6$ $\mu$eV $\mu$m$^2$ and $g_R = 2 g_c$, and condensation rate to $R = 0.01$ ps$^{-1}$ $\mu$m$^2$.

\begin{figure}
\centering

\includegraphics[width=\linewidth]{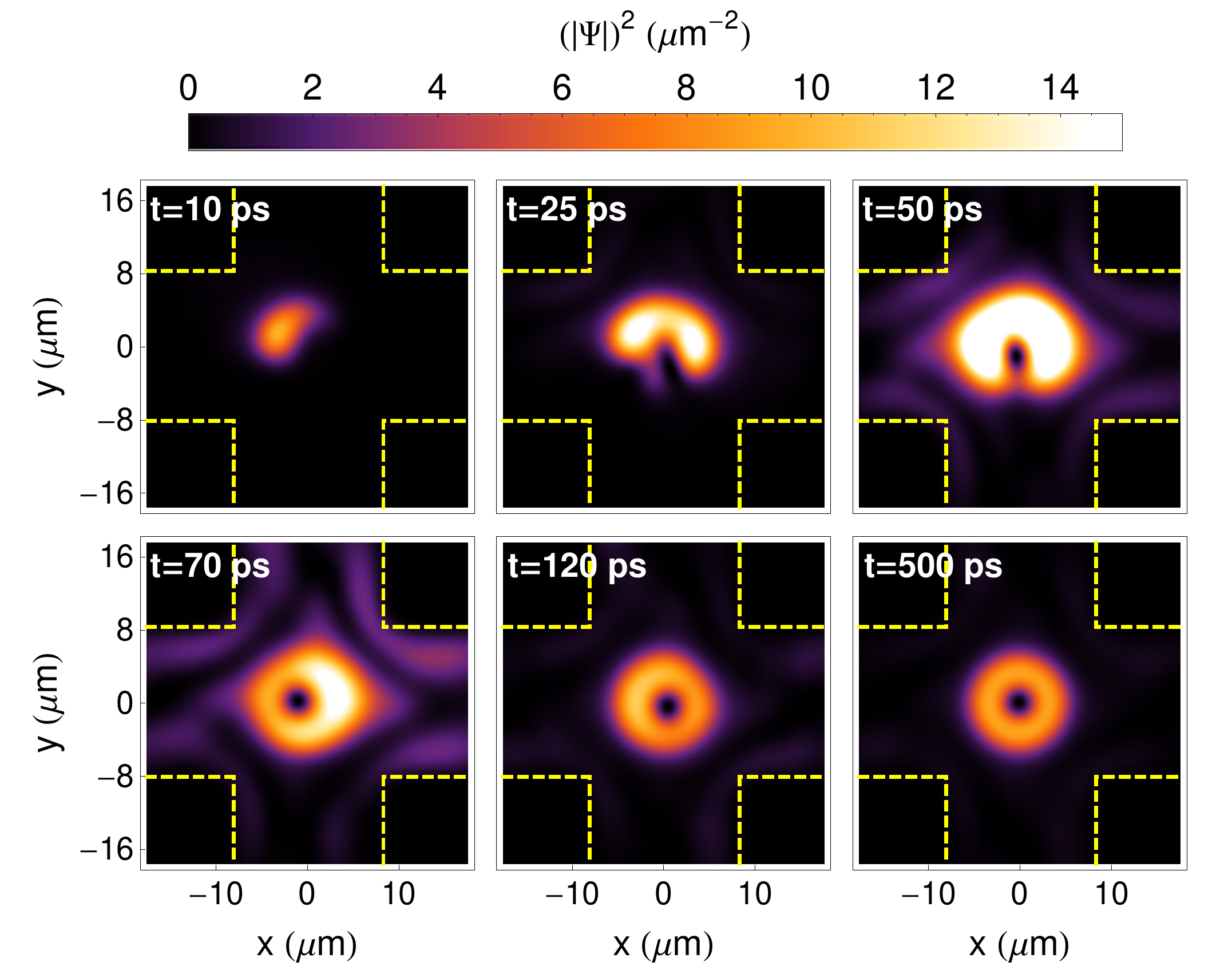}
\includegraphics[width=\linewidth]{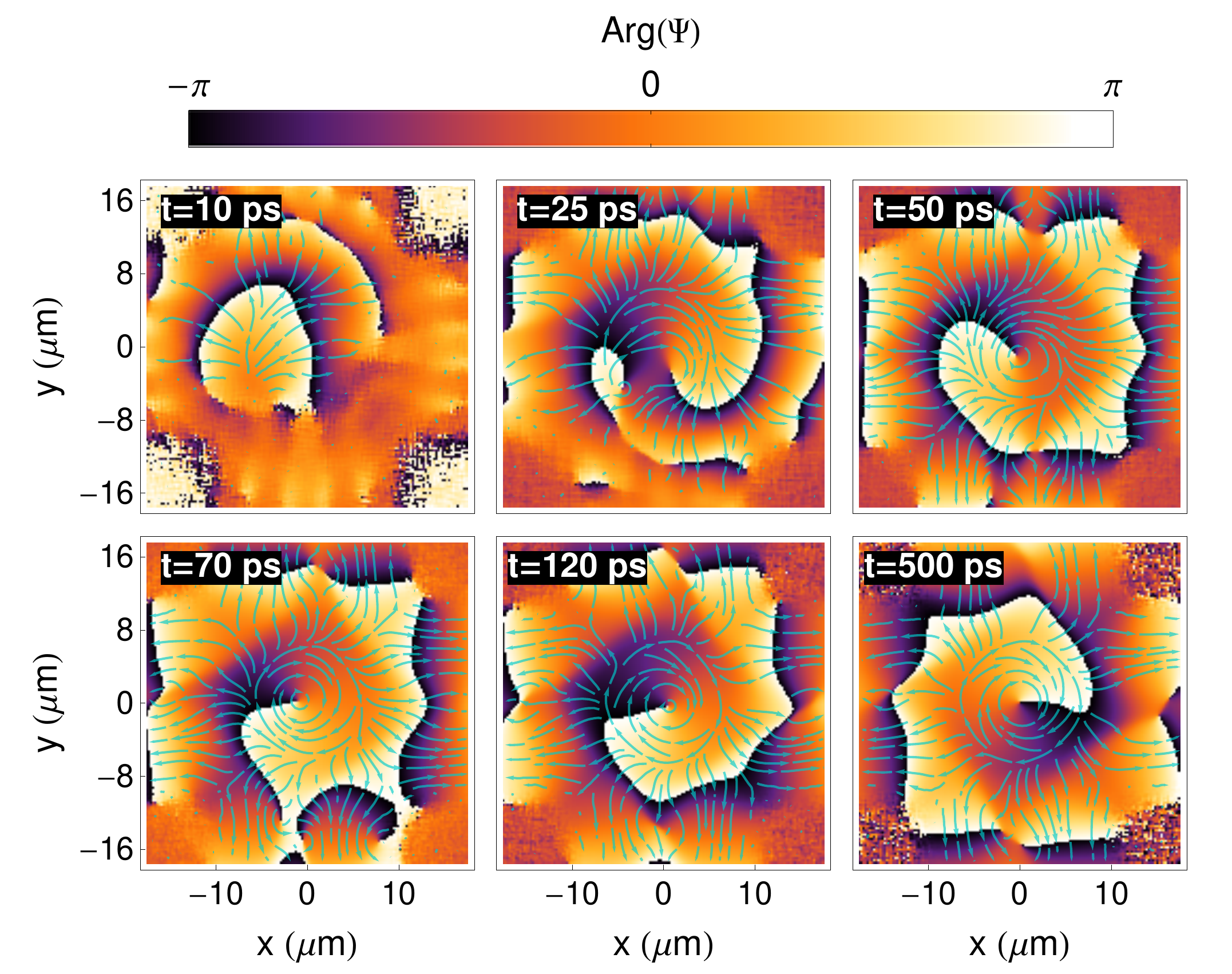}

\caption{(Color online) Polariton density and phase profiles respectively showing the generation of a vortex with charge $m = -1$ by coherent pumping at $\mathbf{r}_c = -3.8\mathbf{\hat{x}}$ $\mu$m in a potential grid node with $\mathbf{k}_c = 0.46\mathbf{\hat{y}}$ $\mu$m$^{-1}$. Edges of the potential guides are outlined by yellow dashed lines. Both the coherent pump and the incoherent ring pump are activated at the same time and after 50 ps the coherent pump is shut off. At 500 ps the polaritons have formed a stable vortex state. Polariton streamlines are plotted along with the phase profiles (blue arrows). Trails of polaritons can be clearly seen as they diffuse away along the guides. The potential grid bulk energy is set to 1 meV, and $P_{R0} = 1$ ps$^{-1}$ $\mu$m$^{-2}$.}
\label{fig3}
\end{figure}

All results displayed in this paper were calculated with different realizations of stochastic noise both as an initial condition and as a weak background noise~\cite{Wouters2008} in order to test the robustness of our results. We start by creating a vortex with either charge $1$ or $-1$. Under an incoherent pump with no coherent pulse, the vortex charge is chosen spontaneously. In order to write vortex states with a definite sign, an additional coherent pulse is applied. This can be accomplished with a Gaussian laser applied within the ring pump area, incident at an angle to the quantum well plane,
\begin{equation} \label{eq.cohpump}
P_c = P_{c0} \exp{\left[-\left(\frac{\mathbf{r}-\mathbf{r}_c}{w_c}\right)^2 - i \left(\frac{E_c t}{\hbar} - \mathbf{k}_c \cdot \mathbf{r} \right) \right]}.
\end{equation}
For the results displayed in Fig.~\ref{fig3} the pump energy is $E_c = 0.18$ meV, the pump width $w_c = 3$ $\mu$m, the momentum $|\mathbf{k}_c| = 0.46$ $\mu$m$^{-1}$, and the amplitude $P_{c0} = 0.3$ meV $\mu$m$^{-1}$. In order to start the circulating flow of polaritons, the coherent pump is placed a small distance $\mathbf{r}_c$ away from the center of the ring pump (yet staying inside the ring) and a clockwise or anticlockwise circulation is created by setting $\mathbf{k}_c$ perpendicular to $\mathbf{r}_c$. The results of creating a stationary vortex state of charge $m=-1$ can be seen stepwise in Fig.~\ref{fig3}. Starting both the coherent- and incoherent pump at the same time, we observe a quick injection of polaritons into the system that starts to rotate. After 50 ps we shut off the coherent pump and allow the system to reach a stationary state. After a few hundred picoseconds the vortex becomes cylindrically symmetric and stable, maintaining its form for as long as the continuous wave ring-shaped pump is applied. In the next sub-section, we demonstrate that the vortex states can be initialized with this procedure using a range of different pump wavevectors, powers, positions and energies.

\subsection{Dependance on coherent pump parameters}
To characterize the process of deterministic vortex formation and study the dependence on the parameters $\{\mathbf{r}_c, \mathbf{k}_c, E_c, P_{c0}\}$, let us introduce the fidelity $F(\Psi,\Psi_0)$ of the vortex state created, $\Psi(\mathbf{r},t)$:
\begin{figure*}
\centering
\includegraphics[width=\linewidth]{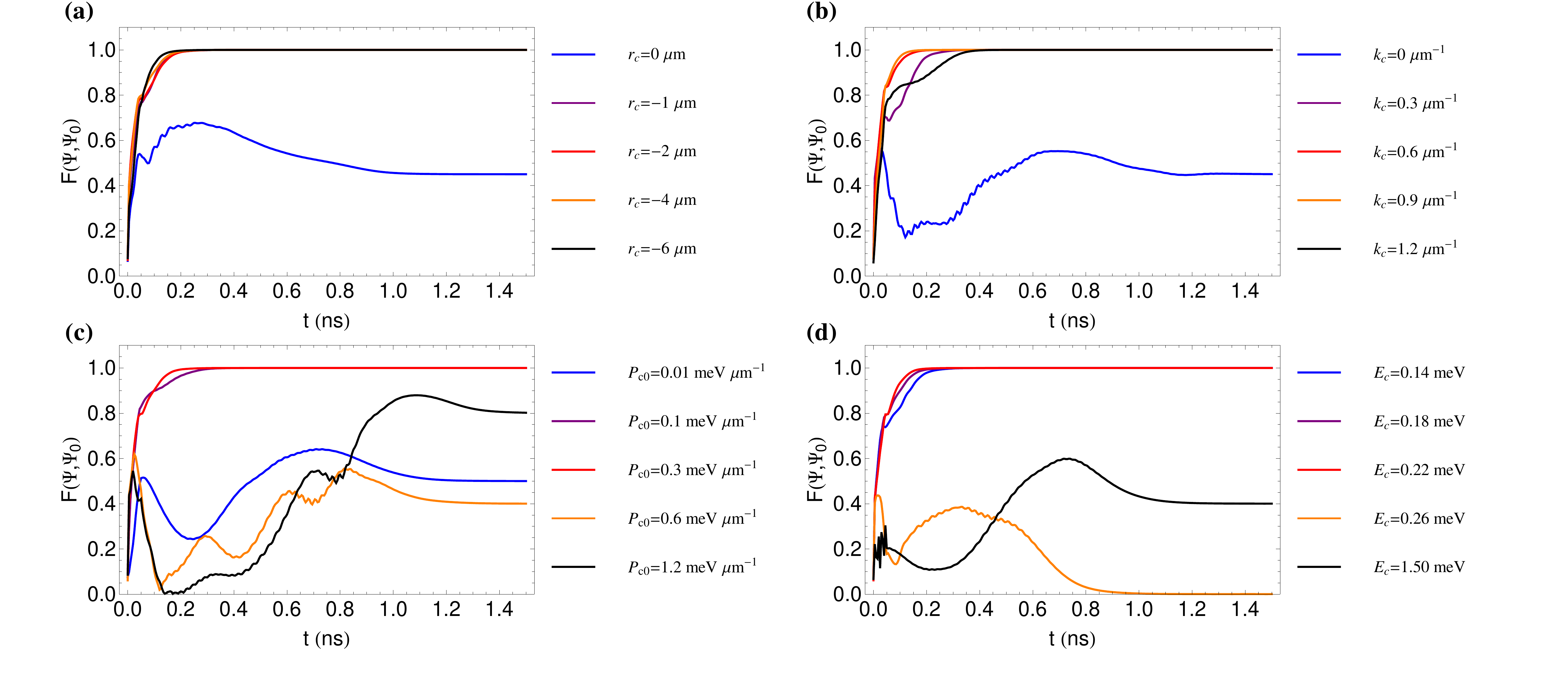}
\caption{(Color online)  Using the same method as in Fig.~\ref{fig3} (excluding the potential grid) we calculate the fidelity $F(\Psi,\Psi_0)$ (Eq.~\ref{eq.fid}) of the state created $\Psi(\mathbf{r},t)$ against a stable vortex state $\Psi_0(\mathbf{r})$ with charge $m = -1$ over 20 realizations of stochastic noise.
(a) $k_c = 0.46$ $\mu$m$^{-1}$, $P_{c0} = 0.3$ meV $\mu$m$^{-1}$ and $E_c = 0.18$ meV.
(b) $r_c = 3.8$ $\mu$m, $P_{c0} = 0.3$ meV $\mu$m$^{-1}$ and $E_c = 0.18$ meV.
(c) $r_c = 3.8$ $\mu$m, $k_c = 0.46$ $\mu$m$^{-1}$ and $E_c = 0.18$ meV.
(d) $r_c = 3.8$ $\mu$m, $k_c = 0.46$ $\mu$m$^{-1}$ and $P_{c0} = 0.3$ meV $\mu$m$^{-1}$.}
\label{fig4}
\end{figure*}
\begin{equation} \label{eq.fid}
F(\Psi,\Psi_0) =\left\langle \frac{\left| \int \Psi_0^*(\mathbf{r}) \Psi(\mathbf{r},t) d\mathbf{r} \right|}{\sqrt{ \int |\Psi_0(\mathbf{r})|^2  d\mathbf{r} \int |\Psi(\mathbf{r},t)|^2  d\mathbf{r} }} \right\rangle,
\end{equation}
where $\Psi_0(\mathbf{r})$ is the target vortex state with charge $m = -1$.  When $\Psi(\mathbf{r},t)$ reaches the same state as $\Psi_0(\mathbf{r})$ then $F(\Psi,\Psi_0) = 1$.

In Fig.~\ref{fig4}a the results of shifting the coherent pump center $\mathbf{r}_c$ along the $x$-axis of the system (the incoherent ring pump being centered at the origin) are shown. Different shifts reveal that that the vortex stabilizes at the same rate. However, as one shifts closer to the center of the ring pump $(\mathbf{r}_c \to 0)$ the circulation induced is no longer definite since the tangential momentum of the polaritons injected decreases (blue line).

Different wavevector magnitudes, where $\mathbf{k}_c$ is oriented along the positive $y$-axis, have a small effect on the vortex created (see Fig.~\ref{fig4}b). A notable difference is for smaller wavevectors (e.g., $|\mathbf{k}_c| = 0.3$ $\mu$m$^{-1}$) and large wavevectors (e.g., $|\mathbf{k}_c| = 1.2$ $\mu$m$^{-1}$) where longer times, about 100-200 ps more, are needed to reach a steady vortex state. If the wavevector is set to zero, then the charge created is random (blue line), corresponding to a fidelity of $0.5$.

We also investigated the dependence on the coherent pumping strength, $P_{c0}$, in Fig.~\ref{fig4}c. When the pumping strength is weak (blue line) then polaritons generated incoherently from the exciton reservoir will overcome the coherent injection, leading to a slower formation of a random vortex state. When the pump is too strong (orange and black line), the polaritons will overcome the potential set by the incoherent ring pump and diffuse away, making the process unreliable.

Finally, in Fig.~\ref{fig4}d the coherent pump energy, $E_c$, reveals that the stability of the vortex created is challenged if one deviates too far away from the polariton resonance energy and a vortex of a random charge is created (black line). Near the regime of resonance, it's possible to have vortices with a definite charge but not necessarily the one aimed for (orange line). When close to resonance (blue, purple and red lines) the choice of a vortex charge becomes definite.

Let us note that when the fidelity is shown as being unity in Fig.~\ref{fig4}, this value is obtained from averaging over a finite number of noise realizations. We can not rule out the possibility of very rare events that may reduce slightly the fidelity in the limit of a very large set of repetitions. Still, in any case, we can conclude that the fidelity is very close to unity from finite numbers of calculations.

\subsection{$2\pi/3$ and $\pi$ rotational symmetric guide setups}
\begin{figure}
\centering
\includegraphics[width=\linewidth]{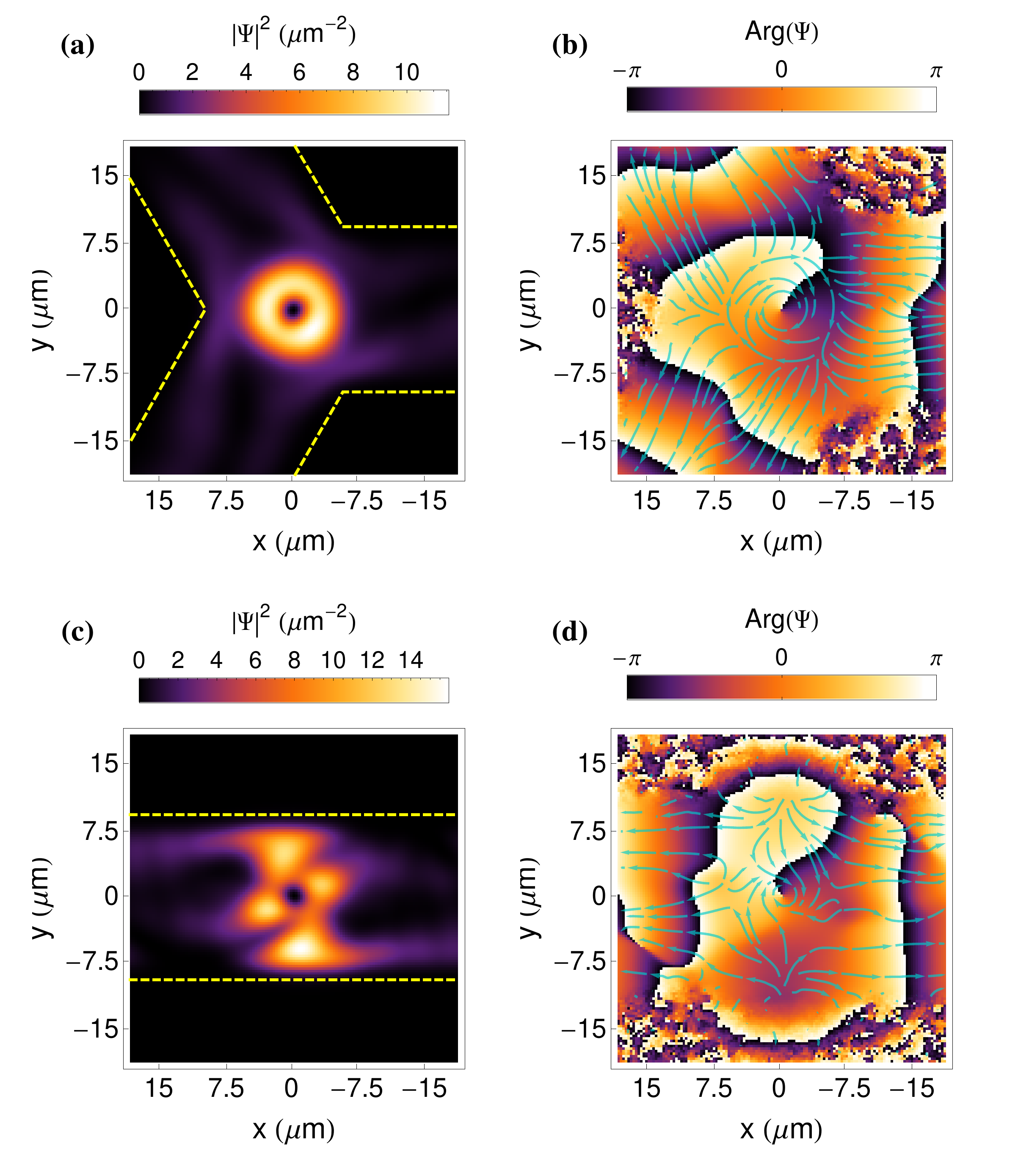}
\caption{(Color online) (a,b) Density and phase profiles of a stable $m = -1$ vortex created in a $2\pi/3$ rotationally symmetric guide setup. (c,d) Density and phase profiles of an unstable $m = -1$ vortex created in a $\pi$ rotationally symmetric guide setup. Blue arrows in (b,d) show the polariton streamlines and yellow dashed lines in (a,c) outline the potential guide edge.}
\label{fig5}
\end{figure}

It was shown in Fig.~\ref{fig3} that a vortex created in a $\pi/2$ rotationally symmetric guide scheme (4 guides leading away from the vortex) becomes stable and cylindrically symmetric after a few hundred picoseconds. However, if the confinement of the vortex is increased, there is an expected increase in polaritons scattering off the guide walls which affects the stability of the vortex. In Fig.~\ref{fig5} we show that a $2\pi/3$ rotationally symmetric guide setup (a,b) supports stationary and cylindrically symmetric vortices whereas in a $\pi$ rotationally symmetric guide setup (c,d) their density profile is non-stationary, indicating an unstable vortex state.

Though the vortex density dip and phase singularity can still be observed in Fig.~\ref{fig5}(c,d), the overall density profile of the state has become severely deformed and any attempts at using it for transferring charge information results in a random charge transferred. Thus $\pi$ symmetric setups are not favorable for vortex control.
\begin{figure}
\centering
\includegraphics[width=0.9\linewidth]{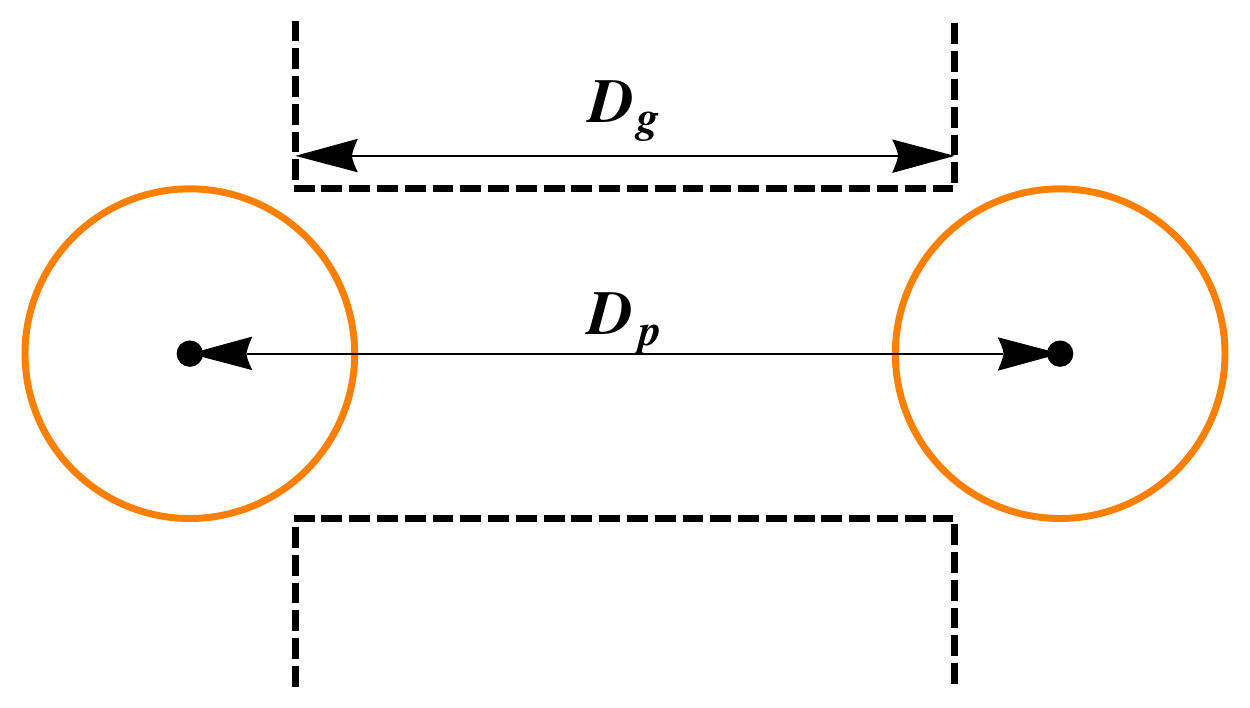}
\caption{(Color online) Schematic of the potential guide edges (dashed lines) and ring pump edges (orange circles) when information transfer takes place. The guide length is defined by $D_g$ and the distance between the ring pump centers is $D_p$. The width of the guide is fixed at 15 $\mu$m.}
\label{fig6}
\end{figure}
\begin{figure}
\centering

\includegraphics[width=\linewidth]{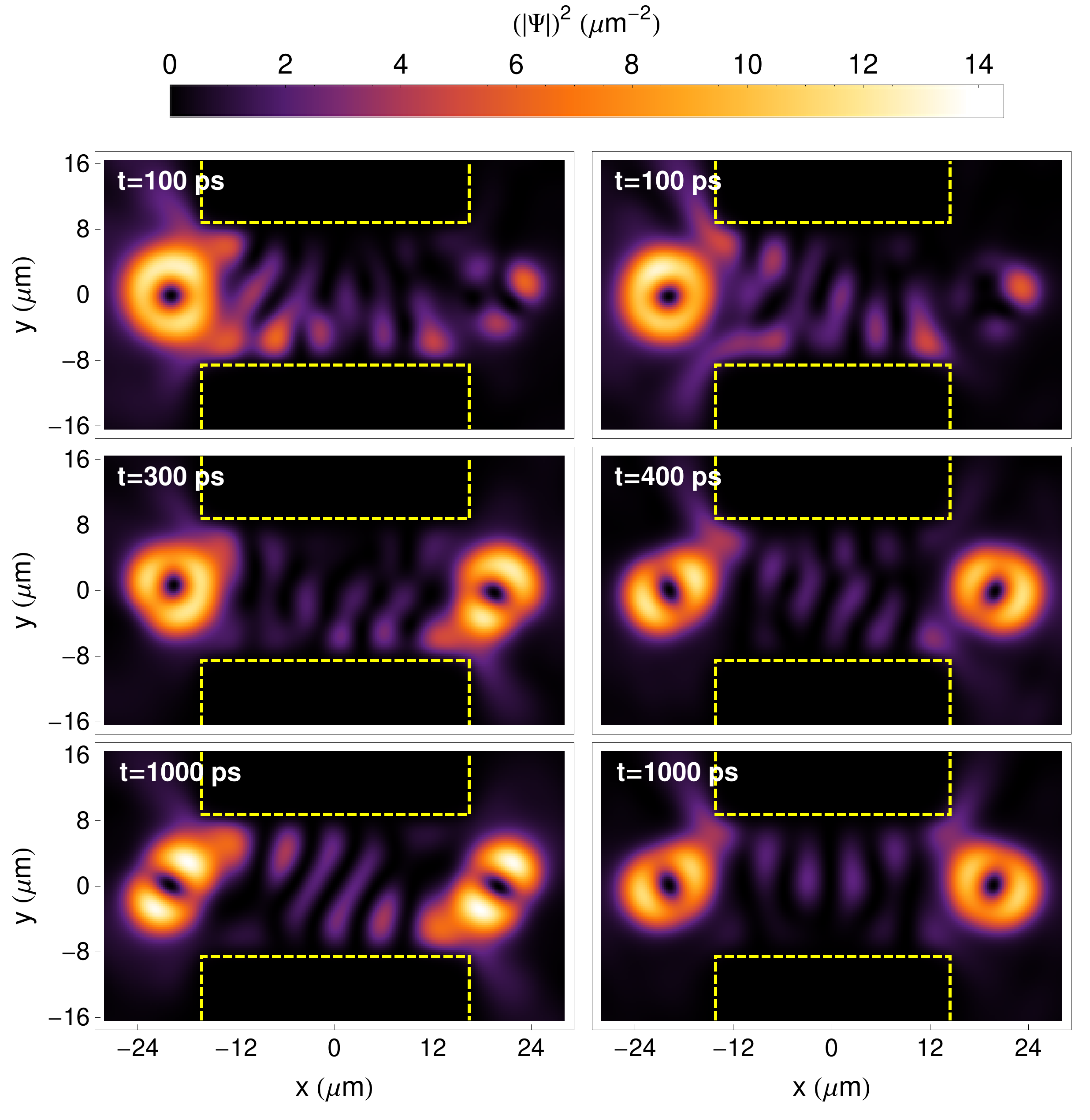}
\includegraphics[width=\linewidth]{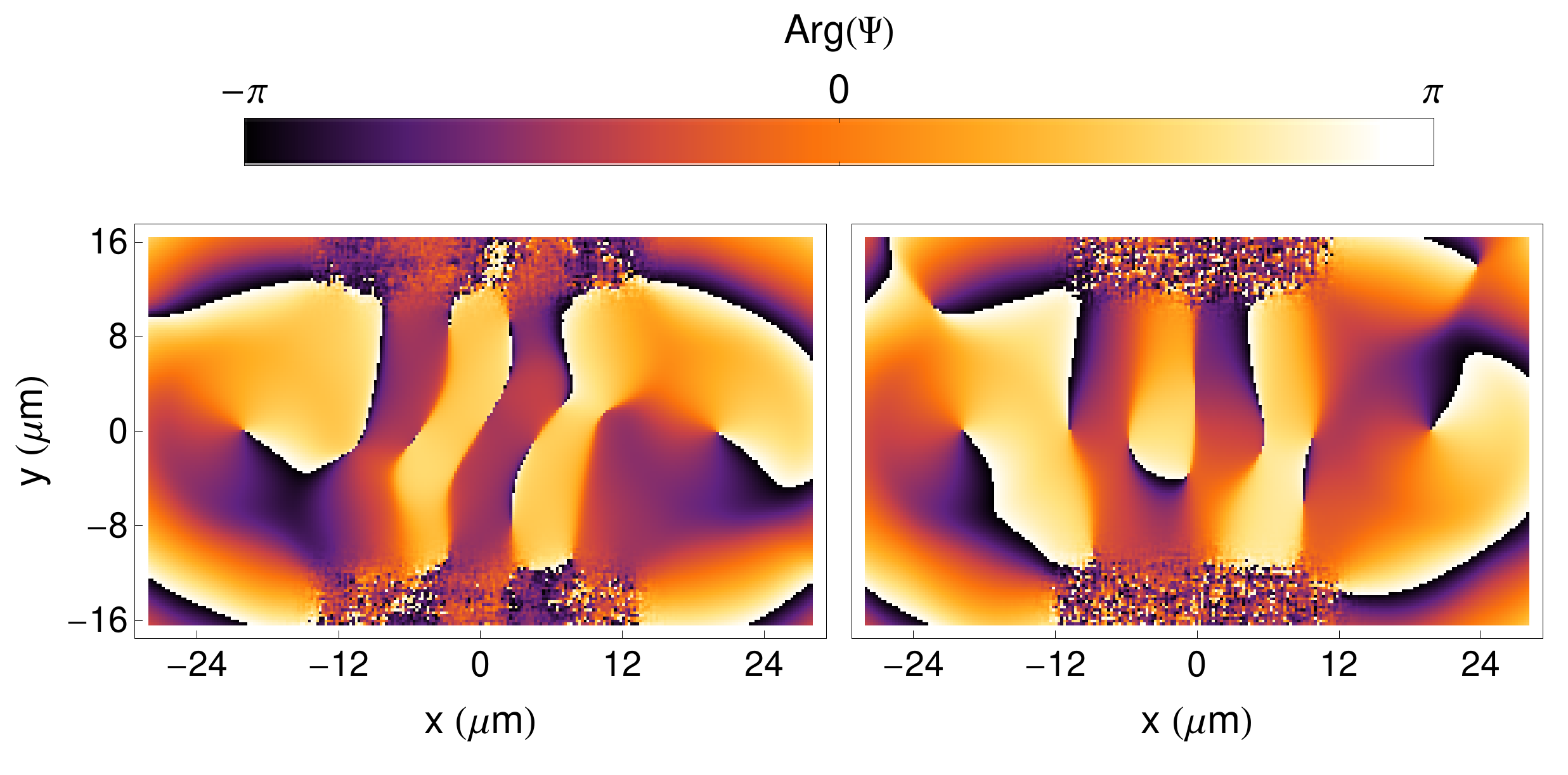}

\caption{(Color online) Left column: Density plots of the copier process taking place at different times. Yellow dashed lines show the edges of the guide. At $t = 300$ ps the transfer is complete and at $t = 1000$ ps the state is nearly stationary. Right column: The inverter process taking place at different times. At $t = 400$ ps the transfer is complete and at $t = 1000$ ps the state is nearly stationary. Bottom panels show the phase profiles at $t = 1000$ ps.}
\label{fig7}
\end{figure}

\section{Operations with vortex states}
We show that by using simple potential guides of different length scales to manipulate the polariton flow pattern, it's possible to copy the same (or inverted) vortex state by activating a second spatially separate incoherent ring pump. Furthermore, if the blue shift of the potential grid is substantially changed then so is the polariton flow pattern in the guides, opening the possibility of controlling the information transfer by having different types of metallic layers on the microcavity. The distance between the two pumps must be chosen such that the vortices can interfere accordingly with each other. If the distance is too small, then polaritons from each pump interfere strongly and the vortex states are lost. If the distance is too great, random noise will overcome the polaritons traveling in the guide.

Fixing the width of the guide at 15 $\mu$m, the two relevant length scales are the lateral length of the guide between the two ring pumps, $D_g$, and the distance between the two ring pump centers, $D_p$ as shown in Fig.~\ref{fig6}. Changing these length scales dramatically affects the diffracted flow pattern of polaritons which can then favor one process above the other even under a large amount of stochastic noise. Note that the ring pump centers are placed equidistant from the center of the guide.

We now describe the method of copying the same or inverted vortex state from one pump to another. The potential guide is set to the desired length scales $(D_p, D_g)$ and on one side a ring pump is activated with a vortex state of either $m = \pm1$ chosen by coherent pumping as shown in Fig.~\ref{fig3}. As the initial vortex settles and becomes stable after 500 ps, we activate a second ring pump with strong random polariton noise in its center as an initial condition. If it were alone, the second ring pump would develop into a vortex state with sign chosen spontaneously as polaritons condense. However, polaritons traveling from the first vortex state arrive at the second with a definite momentum, which depends on the sign of the first vortex state. As in the case of writing the vortex state with a coherent pulse, these polaritons introduce a preferential direction of flow at the position of the second ring pump, which overcomes the strong polariton noise introduced to the system. This allows the second vortex state to form in a way logically dependent on the state of the first.

Results are presented in Fig.~\ref{fig7} where the transfer of charge is shown stepwise in time for two different cases. The yellow dashed lines outline the edges of the guides. Setting $D_g = 30$ $\mu$m and $D_p = 40$ $\mu$m, we observe the formation of a vortex in the second pump with an inverted charge with respect to the charge of the initial vortex (see Fig.~\ref{fig7}, right column). We call this process an \emph{inverter}, $m_1 = \pm1 \to m_2 = \mp 1$.  In the left column of Fig. \ref{fig7}, we observe, for $D_g = 35$ $\mu$m and $D_p = 40$ $\mu$m, the formation of a vortex in the second pump with the same charge with respect to the charge of the initial vortex.  We call this process a \emph{copier}, $m_1 = \pm1 \to m_2 = \pm 1$. After 1 ns the system in both cases has become almost stationary. One can see that in both cases the vortex states become slightly deformed in a dipole manner corresponding to the flow pattern of the polaritons between the pumps.

We expect that faster transfer times can be achieved in microcavities with shorter polariton lifetime or lighter polariton effective mass. Using these results, we can start with any vortex state in one node and transfer either the inverted or same state to any other node in the grid by controlling the distance between the pump centers at each guide. Other grid symmetries, e.g., hexagonal, are also feasible (see Fig.~\ref{fig5}(a,b)).
\begin{figure}
\centering
\includegraphics[width=\linewidth]{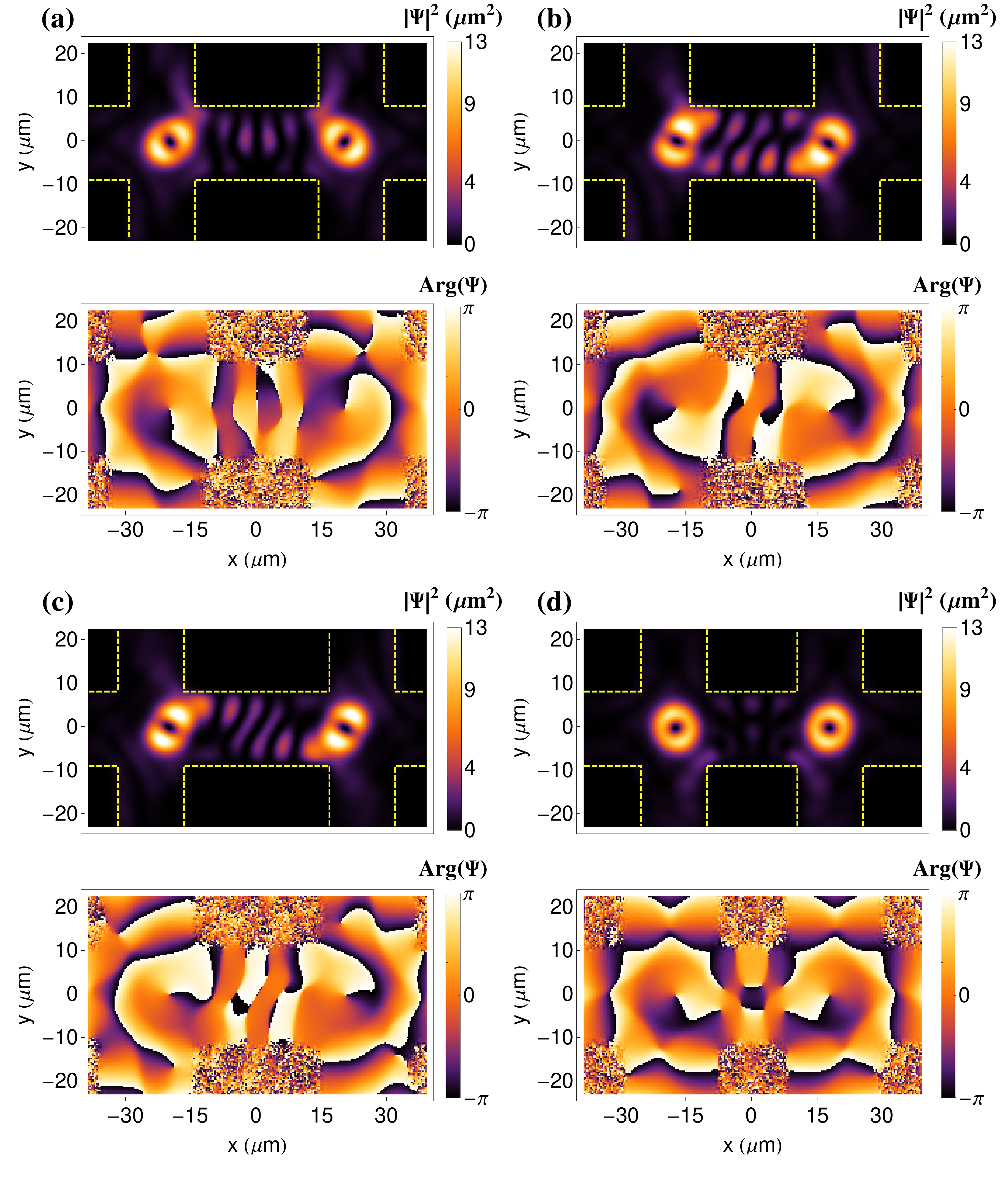}
\includegraphics[width=\linewidth]{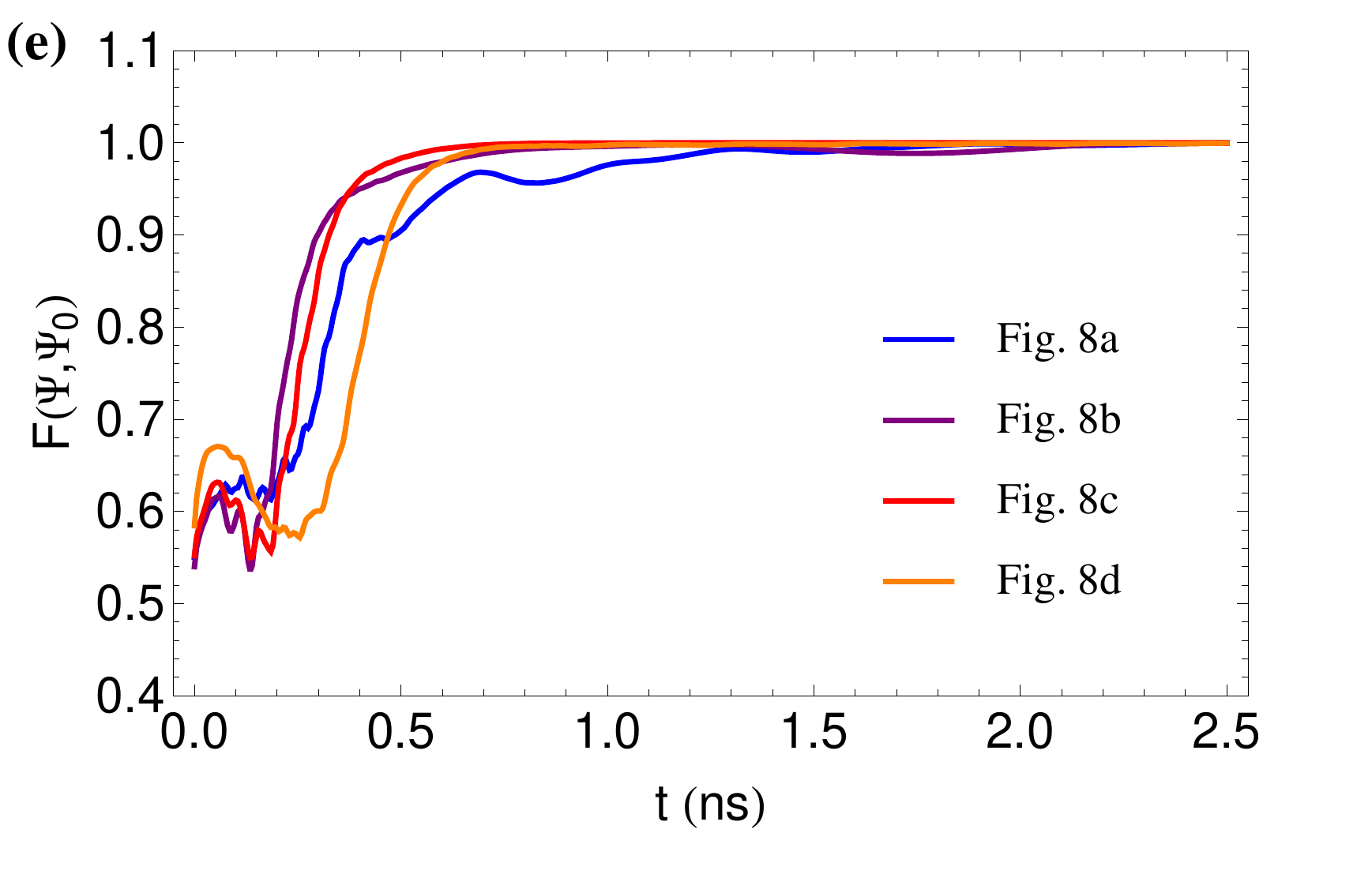}
\caption{(Color online) Four different setups which show a completed transfer of charge information after $2.5$ ns for the copier (b,c) and the inverter process (a,d). Yellow dashed lines outline the edges of the potential grid. (a) $D_p = 40$ $\mu$m and $D_g = 30$ $\mu$m. (b) $D_p = 35$ $\mu$m and $D_g = 30$ $\mu$m. (c) $D_p = 40$ $\mu$m and $D_g = 35$ $\mu$m. (d) $D_p = 35$ $\mu$m and $D_g = 22.5$ $\mu$m. (e) Fidelity shown for each process calculated over 20 realizations of stochastic noise.}
\label{fig8}
\end{figure}

A total of four robust copier and inverter processes were uncovered by varying the length scales $D_p$ and $D_g$ (see Fig.~\ref{fig8}). For simplicity, we restrict to these four cases, while these processes are expected over a wide range of guide parameters. The copier and inverter in Fig.~\ref{fig7} are shown again in Fig.~\ref{fig8}(c,a) respectively but this time at 2.5 ns where the system has become completely stationary. Fig.~\ref{fig8}b shows a copier with $D_p = 35$ $\mu$m and $D_g = 30$ $\mu$m, and Fig.~\ref{fig8}d an inverter with $D_p = 35$ $\mu$m and $D_g = 22.5$ $\mu$m. In order to confirm that these processes are robust we plot the fidelity (Eq.~\ref{eq.fid}) of each case in Fig.~\ref{fig8}e. Each line is calculated for twenty different realizations of stochastic noise. The results show the lines for each process converging to unity indicating that the transfer of charge aimed for takes place with 100\% fidelity within our simulations.

We believe that the four different processes shown in Fig.~\ref{fig8}, along with $\pi/2$ and $2\pi/3$ rotationally symmetric setups shown in Fig.~\ref{fig3} and Fig.~\ref{fig5}(a,b) respectively, can offer many different possibilities in creating potential guide systems in order to efficiently manipulate the transfer of polariton vortex charges. For simplicity, we have neglected the spin degree of polaritons. We expect that the half-vortices that form in spinor polariton condensates~\cite{Rubo2007, Solano2014, Lagoudakis2009} can offer a wider alphabet for topologically protected spin based logic. Future work should focus on the adaptation of vortex bits for use in cascadable logical circuits.~\cite{Ballarini2013,EspinosaOrtega2013}

\section{Conclusion}
We have shown that it's possible to sustain a stable vortex state of charge $m = \pm1$ in an open-dissipative system of exciton polaritons using an incoherent ring shaped pump. The charge of the vortex state can be initialized using a coherent Gaussian pump (within a wide range of pump parameters). These vortex states can furthermore be copied to a different ring pump using simple potential guides. The choice of copying the same charge or the inverted charge can be controlled by either adjusting the lateral length of the guides or the distance between the pumps. Calculating the fidelity of these processes over many realizations of stochastic noise confirmed that they are robust. We believe that a new area of future mechanism can be started using these basic \emph{inverter} and \emph{copier} schemes as building blocks.

{ \it Acknowledgements.---} We thank E. Ostrovskaya and O. Kyriienko for encouraging discussions. This work was supported by FP7 IRSES project "POLAPHEN", FP7 ITN project "NOTEDEV" and Tier 1 project "Novel polaritonic devices".

\end{document}